# INSIDE THE MIND OF A STOCK MARKET CRASH

Stefano Giglio*    Matteo Maggiori†    Johannes Stroebel‡    Stephen Utkus§

First draft: April 4th 2020

This draft: May 21, 2020

**Abstract**

We analyze how investor expectations about economic growth and stock returns changed during the February-March 2020 stock market crash induced by the COVID-19 pandemic, as well as during the subsequent partial stock market recovery. We surveyed retail investors who are clients of Vanguard at three points in time: (i) on February 11-12, around the all-time stock market high, (ii) on March 11-12, after the stock market had collapsed by over 20%, and (iii) on April 16-17, after the market had rallied 25% from its lowest point. Following the crash, the average investor turned more pessimistic about the short-run performance of both the stock market and the real economy. Investors also perceived higher probabilities of both further extreme stock market declines and large declines in short-run real economic activity. In contrast, investor expectations about long-run (10-year) economic and stock market outcomes remained largely unchanged, and, if anything, improved. Disagreement among investors about economic and stock market outcomes also increased substantially following the stock market crash, with the disagreement persisting through the partial market recovery. Those respondents who were the most optimistic in February saw the largest decline in expectations, and sold the most equity. Those respondents who were the most pessimistic in February largely left their portfolios unchanged during and after the crash.

*Yale School of Management, NBER, CEPR: stefano.giglio@yale.edu.
†Graduate School of Business, Stanford University, NBER, CEPR: maggiori@stanford.edu.
‡Stern School of Business, New York University, NBER, CEPR: johannes.stroebel@stern.nyu.edu.
§Vanguard: steve_utkus@vanguard.com.

Stephen Utkus is employed at Vanguard in a research capacity. Giglio, Maggiori, and Stroebel are unpaid consultants at Vanguard in order to access the anonymized data. Vanguard provided anonymized portfolio and survey data as well as survey research services for this project. We thank Jose Scheinkman and Alp Simsek as well as seminar participants at Stanford for helpful comments. The authors would also like to thank Catherine Clinton, Sophia Bunyaraksh, and Jean Young at Vanguard for their assistance with the survey and with understanding the data.

The dynamics of beliefs play a central role in macroeconomics and finance. As a result, analyzing how beliefs vary with changes in the economic environment and how they affect investment choices is important for understanding asset markets and real economic activity. This paper offers a unique picture of the joint dynamics of beliefs and trading during a time of severe market distress, for a relevant set of market participants: retail investors who are clients of Vanguard, one of the world's largest asset managers. To conduct this analysis, we field a large-scale survey to elicit Vanguard investors' beliefs about future economic growth and stock market returns before, during, and after the stock market crash of March 2020, which was triggered by the COVID-19 pandemic. We then describe the evolution of beliefs and disagreement, as well as the relationship between beliefs and investors' trading activity during this period.

This paper is part of an ongoing project that we launched in 2017 in collaboration with Vanguard, with the aim of deepening our understanding of expectations in macroeconomics and finance, and to provide insight into the relationship between beliefs and portfolio decisions (see Giglio et al., 2019, for details). The heart of this project is a newly-designed survey, the GMSU-Vanguard survey, that elicits beliefs central to macro-finance. These beliefs include expected stock returns and expected GDP growth in both the short run and the long run, as well as respondents' perceived probabilities of economic and stock market disasters. In this paper, we explore three waves of the survey from early 2020. The first wave was administered in mid-February, near the peak of the U.S. stock market; the second wave in mid-March, after the U.S. stock market had declined by about 20% from its peak; and the third wave was administered in mid-April, after the stock market had rallied by 25% from its low point, though it was still about 17% below its peak.

We find that average beliefs about stock returns over the next year turned substantially more pessimistic following the stock market crash; average expectations of GDP growth over the short-term (the next 3 years) also declined, although only moderately. Average expectations of short-run disaster probabilities in stock returns and GDP growth, defined respectively as a stock market drop of 30% or more in the next year and annual real GDP growth of less than -3% over the next 3 years, both spiked during this episode. On the other hand, long-term expectations of GDP growth and stock returns over the next 10 years remained stable or even increased somewhat.

The dispersion of beliefs across investors, often referred to as disagreement, increased substantially during the crash. Interestingly, the beliefs of optimists and pessimists, classified according to their pre-crash beliefs, moved in substantially different ways during this period. Overall, the vast majority of investors became more pessimistic about the short-run outlook of the stock market. However, among those investors who were relatively pessimistic before the crash (i.e., those who, in February, were expecting negative 1-year stock market returns), about half actually revised their expectations upwards in the March and April survey waves.

An important feature of our study is the ability to match survey responses to the clients' portfolios and daily trading activity at Vanguard. This allows us to conduct a "high-frequency" study of the relationship between beliefs and investment decisions at the individual level. February and March 2020 were periods of elevated trading activity both at Vanguard and more generally in the markets. Consistent with the findings of Giglio et al. (2019), we show that, before the crash, respondents who were more optimistic about stock market returns had a higher fraction of their portfolio invested in equity. We then document that when the crash occurs at the end of February, Vanguard clients in our sample rebalance their portfolios away from equities. Those investors who were ex-ante more optimistic sell more equity immediately after the crash. Those investors who were initially more pessimistic keep their portfolios largely unchanged. The trading decisions, therefore, align closely with the differential belief dynamics for initial optimists and pessimists.

The aim of this paper is to document these patterns in the data. We take no stance on whether the expectations measured by our survey are rational or include behavioral elements, or whether the trading decisions that we document were optimal. However, we note that even though the dynamics of individual and aggregate expectations after large shocks are among the most informative moments for models of macroeconomics and finance, they are rarely observed.[1] This is both because the events themselves are rare and because large-scale surveys that track people over time have only recently become available. Therefore, we believe that our findings have the potential to contribute a new important moment that can be used to calibrate and evaluate different models, especially models that feature rare disasters or belief heterogeneity. In Section III we review the main qualitative implications for such models.

## I  Brief Survey Description

We explore responses to three waves of the GMSU-Vanguard survey. This survey elicits the beliefs of Vanguard investors about expected stock returns and expected GDP growth in both the short run and the long run, as well as investors' perceived probabilities of economic and stock market disasters. The Appendix presents the full survey flow and the exact wording of the various questions. The survey has been administered to retail and retirement clients of Vanguard every two months since February 2017.[2] The surveyed population is one that is relevant for understanding financial markets: retail investors with substantial investments in both equity and fixed income products. The median respondent has 225,000 USD invested with Vanguard — 70% in equity instruments and 15% in fixed income instruments — and is approximately 60 years of age. Giglio et al. (2019) provides more detailed summary statistics on the investor population sampled by this survey.

As part of our ongoing project, a regular survey was administered on February 11th, 2020, which turned out to be almost exactly the all-time high in the U.S. stock market. At this time, the COVID-19 outbreak in China had already occurred, but its implications had not yet been widely reported or understood. This survey wave therefore offers us a measure of investor beliefs before the subsequent crash. After one of the longest and most pronounced stock market booms on record during 2009-2019, the U.S. stock market then experienced a sudden crash starting on Monday, February 24th. By March 11th, the S&P 500 index had dropped 19.2% from its previous high. On that day, the financial press announced that U.S. stock markets had entered "bear market territory," commonly defined as a drop in value of 20% or more from the high point. Following these dramatic market events, we fielded an unscheduled survey on March 11th, 2020, at 6pm EDT, after the market close.³ After this survey wave, the market fell further, bottoming out at 34% below its peak on March 23, 2020. On April 16th, we fielded another one of the project's regular bi-monthly survey waves. By that date, the stock market had rallied by 25% from its lowest point, though it was still about 17% below its February peak. By this time, newspapers had devoted substantial coverage to the impacts of COVID-19 on the real economy. Figure I shows the dynamics of the S&P 500 index during the this period, as well as the exact timing of our surveys.

---

[1]In the wake of the Covid-19 crisis there is a wave of interesting work measuring expectations with different approaches (see, for example, Gormsen and Koijen, 2020; Landier and Thesmar, 2020).

[2]The sample selection rules are described in Giglio et al. (2019), and we encourage the reader to refer to that paper for more background information on the survey. The only difference to the sample selection approach described in Giglio et al. (2019) is that the March survey wave did not add newly selected clients that had never been contacted before by our study. This is consistent with the focus in this article on changes in beliefs since February.

**Figure I:** S&P 500 and Dates of Expectation Survey

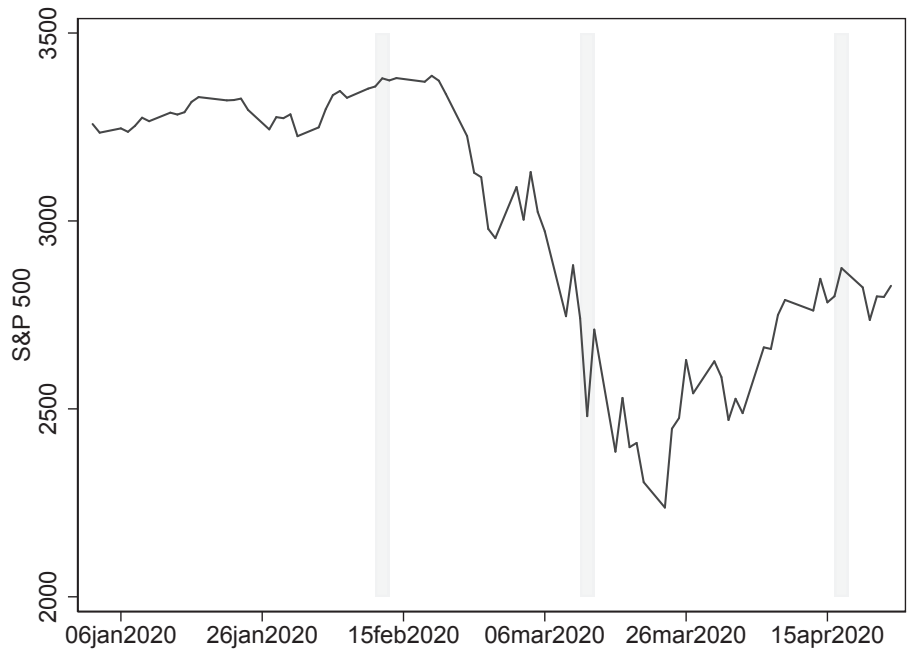

**Note:** Figure shows the development of the S&P 500 index between between January 2, 2020, and April 24, 2020. The three shaded regions correspond to the dates of the February, March, and April waves of the GMSU-Vanguard survey.

Our regular survey obtains approximately 2,000 responses per wave, with the majority of responses coming from people who responded to previous waves. The February wave obtained 2,390 responses. The March wave obtained 1,862 responses, and 484 of those responses came from individuals who had also responded to the February wave.⁴ The April wave obtained 2,414 responses, and 670 of those responses came from investors who responded to the February wave.

---

³When we started the project, we had anticipated that one of the most interesting questions was how beliefs would change during an economic crisis or a large stock market crash. We had therefore designed the administration of the survey to be able to launch additional surveys at short notice following such events. This paper is the outcome of this pre-planned contingent survey administration.

⁴Response rates vary on average between 4% for newly contacted people to above 10% for those who have already responded once. Giglio et al. (2019) discuss various dimensions of selection bias in who answers the survey.

## II  BELIEF DYNAMICS AND TRADING OVER THE COVID-19 CRASH

**Dynamics of Average Beliefs.**   We begin by documenting patterns in average beliefs in the data. Figure II shows the time series of average beliefs across all our respondents for the entire period covered by our survey.[5] Panel A shows the 1-year expected stock market return, Panel B the 10-year expected annual stock market return, Panel C the expected annual real GDP growth over the next three years, Panel D the expected annual real GDP growth over the next ten years, Panel E the probability of stock market disaster (defined as a loss of more than 30% within the next year), and Panel F the probability of a GDP disaster (defined as average annual GDP growth of less than -3% over the next three years).

Many of the panels in Figure II show large changes in beliefs in the two survey waves following the stock market crash, changes of a magnitude not observed in the previous two years. Specifically, in the two years before the crash, expectations about 1-year stock market returns had ranged between 3% and 6%, and were at the high end of that range in February 2020. The crash brought them down to the 1% to 2% range. This pessimism about short-run market returns is not accompanied by pessimism about the long run. Expected annual stock market returns over the next ten years actually increased modestly after the crash, from 6.9% per year to 7.2% per year.

Panel C of Figure II shows that average expectations about real GDP growth over the next 3 years moved from 2.8% to 2.2% following the crash.[6] Similar to the expectations for stock returns, a fall in short-run expectations is associated with an increase in long-run expectations: annual 10-year growth expectations increased from 3.1% to 3.5% per year. To provide a sense of the order of magnitude of expected GDP losses, it is illustrative to compare the expectations to what actually happened during the global financial crisis. Starting at the end of June 2008, real GDP growth in the U.S. over the next three years was 0.3%, with a v-shaped pattern of growth over that period. At least for now, the investors surveyed here are not expecting that the COVID-19 shock will lead to GDP losses as large as those experienced during the financial crisis.[7]

Panels E and F of Figure II show large increases in the perceived probabilities of short-run disasters in stock market returns and GDP growth. The probabilities of such disasters increase from 4.5% to 8% for the stock market, and from 5% to 8.5% for GDP growth. It is these extreme outcomes that the ongoing economic policy response is trying to minimize, but our respondents still find their probability to have increased substantially.

---

[5]Similar to Giglio et al. (2019), we set extreme outlier answers (i.e., responses below the bottom percentile or above the top percentile) for each unbounded expectation question equal to missing. Naturally, there are some critical judgment calls involved in selecting these cutoffs, which involve trading-off retaining true extreme beliefs with excluding answers from individuals who probably misunderstood the question or the units of measurement. Giglio et al. (2019) conducted extensive sensitivity analysis to confirm that our results are robust to a wide range of choices for the cutoff values. We also confirmed that the results are robust to winsorizing extreme answers rather than setting them equal to missing, and to dropping all answers of individuals who report extreme answers to at least one question.

[6]The median expectation also moves down from 2.3% in February to 2% in April, so that the change in the average is not driven by changes in outliers over time.

[7]The short-run economic growth expectations elicited in this survey are also considerably higher than those of professional forecasters. In their World Economic Outlook released in April 2020, the IMF forecasted U.S. GDP growth of -5.9% in 2020 and 4.7% in 2021. The Wall Street Journal survey of professional forecasters in April 2020 reported average expected annual growth for the period 2020-2022 at 0.88%.

**Figure II:** Average Responses to Expectation Survey

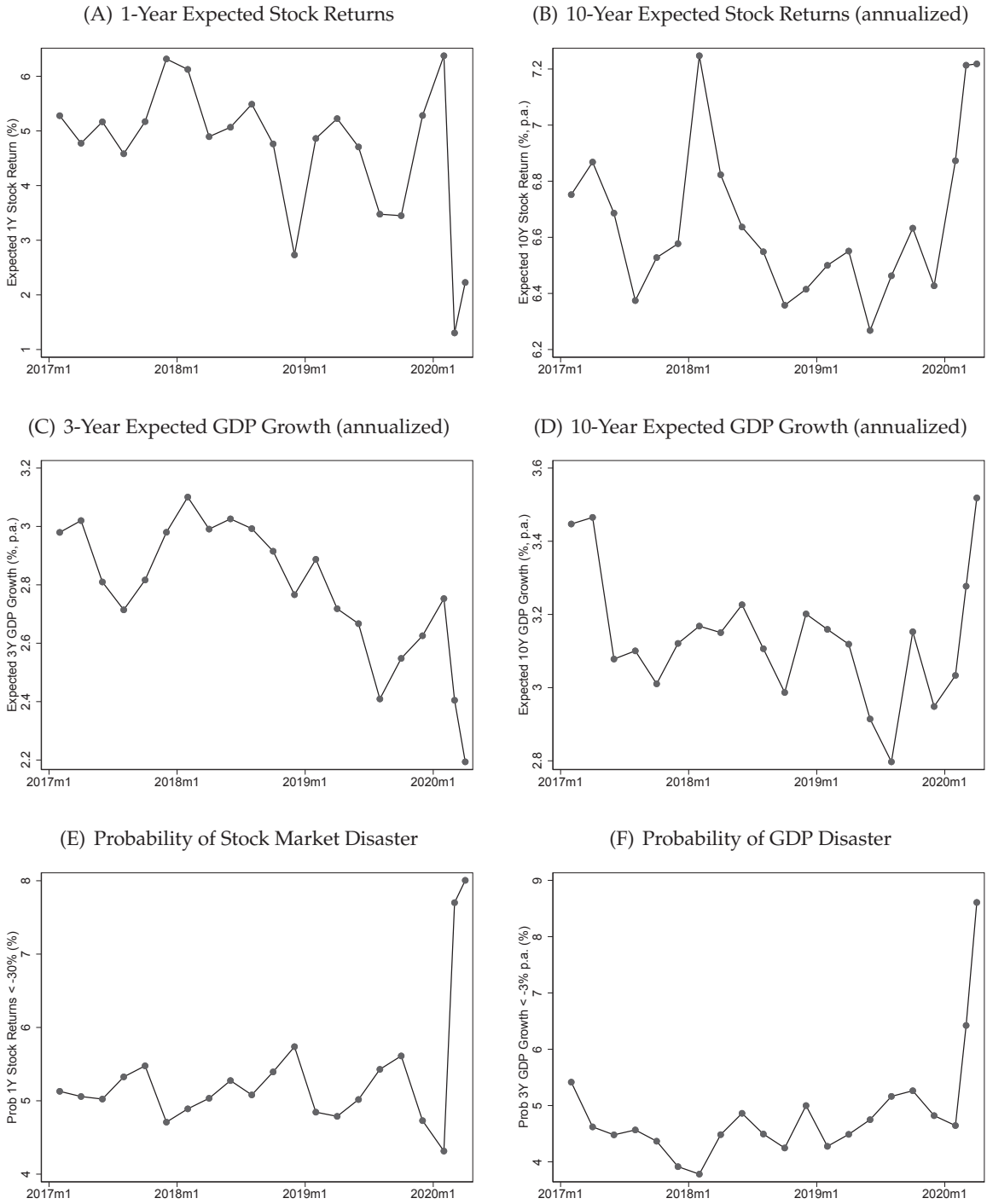

**Note:** Figure shows average beliefs across all respondents in each wave of the GMSU-Vanguard survey. Panel A shows the 1-year expected stock market return, Panel B the 10-year expected stock market return (annualized), Panel C the expected real GDP annual growth over the next 3 years, Panel D the expected real GDP annual growth over the next 10 years, Panel E the probability of stock market returns being lower than -30% over the next year, and Panel F the probability of GDP growth being less than zero on average over the next 3 years.

**Dynamics of Belief Disagreement.** Beyond studying the behavior of average beliefs across investors, our data also allow us to understand the evolution of disagreement among investors. Figure III shows smoothed kernel densities of the cross-section of beliefs, for the 1-year expected return (Panel A) and for the probability of a stock market disaster (Panel B). In each Panel we plot three densities, each corresponding to a different survey wave (February, March, and April).

**Figure III:** Distribution of Responses to Expectation Survey

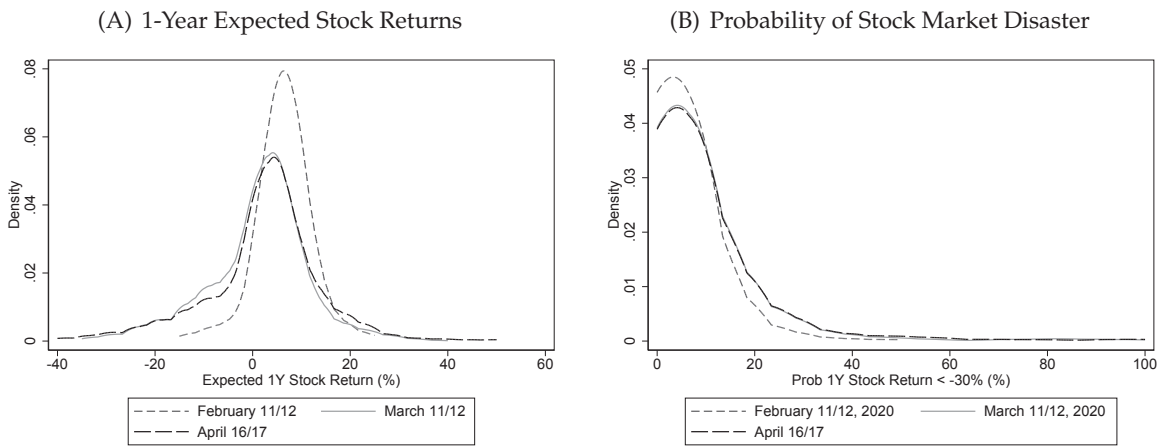

**Note:** Figure shows kernel density distributions over responses on the February 11-12, 2020, the March 11-12, 2020, and the April 16-17, 2020 waves of the GMSU-Vanguard survey. The left panel shows the distribution of beliefs about 1-year expected stock returns, the right panel shows the distribution of beliefs about the probability of a stock market decline of more than 30 percent over the coming 12 months.

The dispersion in beliefs across individuals — the level of disagreement — increased substantially after the market crash, as visible from the fattening of densities. The cross-sectional standard deviation (across respondents) of reported 1-year expected stock returns almost doubled from 5.3% to 10.1% between the February and March waves. The April survey shows a level of disagreement very similar to the March one, despite the fact that the stock market had rallied substantially in the meantime. The response of disagreement is asymmetric, with pessimism — the left tail of Panel A — becoming substantially more pronounced in the investor population: the cross-sectional skewness of beliefs increases from -0.32 to -0.47 from February to March. Consistent with this finding, the 10th percentile of the belief distribution moves from 2% to -10%, whereas the 90th percentile remains essentially stable, falling to 10% from 12%. Finally, the 90th percentile of the distribution of perceived probabilities of disaster, captured by the right tail of Panel B, doubles from 10% to 20% between the February and March waves of the GMSU-Vanguard survey.

We can further refine our understanding of belief dynamics by digging deeper into the nature of disagreement. We seek to understand which people changed their beliefs, and how: did pessimists become more pessimistic, or was the change in disagreement driven largely by investors who were previously optimistic? Our survey is well suited to answering these questions, because we observe a significant number of investors who respond to multiple waves of the survey.

In Table I, we study the subset of investors that responded to the February wave as well as at least one of the March or April waves of the GMSU-Vanguard survey. Panel A focuses on the

## Table I: Changes in Beliefs by Initial Belief

*Panel A.I.*

| Expected 1y Stock Returns (%, Feb) | Change in Expected 1y Stock Returns (Feb - Mar, ppt) | | | | | |
|---|---|---|---|---|---|---|
| | Less than -20 | Between -20 and -10 | Between -10 and -5 | Between -5 and 0 | Between 0 and 5 | Greater than 5 |
| Less than 0 | 0.0% | 3.1% | 9.4% | 25.0% | 18.8% | 43.8% |
| Between 0 and 5 | 8.8% | 17.6% | 10.8% | 35.3% | 17.6% | 9.8% |
| Between 5 and 10 | 10.0% | 13.1% | 12.4% | 41.3% | 13.9% | 9.3% |
| Greater than 10 | 3.8% | 29.5% | 29.5% | 24.4% | 5.1% | 7.7% |

*Panel A.II.*

| Expected 1y Stock Returns (%, Feb) | Change in Expected 1y Stock Returns (Feb - Apr, ppt) | | | | | |
|---|---|---|---|---|---|---|
| | Less than -20 | Between -20 and -10 | Between -10 and -5 | Between -5 and 0 | Between 0 and 5 | Greater than 5 |
| Less than 0 | 0.0% | 6.1% | 14.3% | 6.1% | 20.4% | 53.1% |
| Between 0 and 5 | 8.0% | 13.3% | 8.0% | 30.0% | 27.3% | 13.3% |
| Between 5 and 10 | 5.2% | 12.0% | 15.2% | 37.3% | 17.2% | 13.1% |
| Greater than 10 | 8.5% | 18.6% | 17.8% | 30.5% | 11.0% | 13.6% |

*Panel B.I.*

| Expected Probability of Crash (%, Feb) | Change in Probability of Crash (Feb - Mar, ppt) | | | | | |
|---|---|---|---|---|---|---|
| | Less than -5 | Between -5 and 0 | Between 0 and 5 | Between 5 and 10 | Between 10 and 20 | Greater than 20 |
| Between 0 and 2.5 | 0.0% | 4.9% | 64.8% | 14.8% | 9.8% | 5.7% |
| Between 2.5 and 5 | 0.0% | 50.0% | 21.4% | 14.3% | 14.3% | 0.0% |
| Between 5 and 10 | 1.8% | 23.6% | 42.7% | 19.1% | 9.1% | 3.6% |
| Greater than 10 | 35.9% | 18.5% | 18.5% | 5.4% | 14.1% | 7.6% |

*Panel B.II.*

| Expected Probability of Crash (%, Feb) | Change in Probability of Crash (Feb - Apr, ppt) | | | | | |
|---|---|---|---|---|---|---|
| | Less than -5 | Between -5 and 0 | Between 0 and 5 | Between 5 and 10 | Between 10 and 20 | Greater than 20 |
| Between 0 and 2.5 | 0.0% | 6.3% | 65.4% | 13.2% | 10.8% | 4.2% |
| Between 2.5 and 5 | 0.0% | 25.0% | 39.3% | 17.9% | 10.7% | 7.1% |
| Between 5 and 10 | 0.8% | 25.4% | 36.5% | 22.2% | 11.1% | 4.0% |
| Greater than 10 | 39.4% | 21.2% | 16.7% | 6.1% | 10.6% | 6.1% |

**Note:** Panel A shows the transition density between the level of ex-ante expectations about 1-year stock market returns (rows) and ex-post changes in these expectations (columns). Panel A.I focuses on the transition between February 2020 and March 2020; Panel A.II focuses on the transition between February 2020, and April 2020. Panel B shows an analogous analysis for the perceived probability of the stock market return over the coming year being lower than -30%. The interior buckets in both rows and columns are closed on the left and open on the right.

1-year expected return and Panel B focuses on the probability of a stock market disaster. In Panel A, we group respondents into four buckets based on their beliefs about 1-year stock returns before the crash in the February wave; each row corresponds to a different group. Those investors that, in the February wave, were most pessimistic are shown in the top row; this group expected negative returns going forward. The bottom row, instead, includes the most optimistic investors, those that in February expected 1-year stock returns above 10%. The columns report the change in beliefs (equally weighted) in percentage points between February and March in Panel A.I, and between February and April in Panel A.II. Each entry reports the fraction of investors within each row that experienced a changed in belief in the range expressed in the corresponding column. For example, the first row shows that of those investors that expected negative returns in February, 3% lowered their expectations by 10-20 percentage points; 9% lowered their expectations by 5-10 percentage points; and 44% increased their expected stock returns by more than 5 percentage points.

Panel A of Table I shows a widespread transition toward more negative beliefs across most investors.[8] For example, 87% of the previously-most-optimistic group become more pessimistic in

---

[8]Despite this large time-series decline in expected returns, Fact 3 from Giglio et al. (2019) — that the panel variation in beliefs is best explained by individual fixed effects and not time fixed effects — continues to hold. Focusing on

March. One exception is the group that includes the most pessimistic respondents in February; for this group, 63% of the respondents increase their expectations between February and March, and 73% become more optimistic between February and April. One interpretation of these results is that in February, after the spread of the Coronavirus had already started, a set of individuals (the pessimists) thought a stock market crash was likely to occur over the next year. As this scenario actually unfolded, about half of these individuals thought that stock prices had fallen far enough as to increase their expected returns going forward; the other half expected further stock market declines. On the other hand, the vast majority of optimists revised their expectations downwards in light of the market crash that they did not anticipate in February.

This view is also supported by Panel B of Table I, which presents an analogous analysis for the perceived probability of a stock market disaster (in this panel, initial pessimists are in the last row). Those who ex-ante reported the highest probabilities of a large stock market decline are also those who decreased their perceived probability the most following the actual realization of such a decline: just over half the pessimists become more optimistic.

**Trading Behavior.** We find that both the levels and dynamics of beliefs are reflected in portfolio choice and trading activity. Figure IV shows the dynamics of portfolios over February and March for the respondents to the February survey, grouped by the level of their expected 1-year stock returns in February. We label those respondents who are in the top tercile of the February belief distribution as "optimists," those in the middle tercile as "neutrals," and those in the bottom tercile as "pessimists."[9] The percentage of each individual's portfolio that is invested in equity is strongly associated with her expected stock returns. On January 31st, the date at which we measure the portfolios using market values, optimists have, on average, 73% of their portfolio invested in equity. The average equity percentage is 66% for neutrals and 62% for pessimists. This result is consistent with the findings in Giglio et al. (2019), who documented that individual beliefs are associated with portfolio choice, but also that the relationship is quantitatively more muted than in frictionless benchmark models.

For February and March, we construct portfolios for each respondent keeping prices constant at their January 31st levels. As a result, the portfolio dynamics in Figure IV reflect active portfolio trading, as opposed to changes in market values.[10] Panel A of Figure IV focuses on all respondents and shows that, in accordance with the belief dynamics by group described above, the optimists sell the most equity (on average, they actively decrease their equity share by 1.05%), followed

---

individuals who have responded to at least three survey waves since February 2017, the $R^2$ of a regression of panel beliefs on individual fixed effects is 51.2%, while it is 5.6% for a regression of the panel beliefs on time fixed effects. Similar patterns hold for all beliefs elicited in this survey.

[9]We group in terciles, rather than the finer groups in Table I, to maintain equal sized groups and sufficient statistical power given that most people do not trade. The average expected stock market returns over the next year (February wave) for the three groups are: 12.2% for the optimists, 7% for the neutrals, and 2.1% for the pessimists. Thus the 'pessimists' in this analysis are not as extremely pessimistic as the lowest group in Table I.

[10]One exception is that when respondents trade we value the trade at the actual transaction price. In the context of our study this is likely to be conservative in the sense that it underestimates the change in portfolio allocation. This occurs because when agents sell equity after the crash they do so at lower prices. Similarly, the share of portfolio equity at market value is falling during this period also for those who do not trade. While we focus on active trading, we also stress that "not trading" to rebalance a portfolio after market changes is also an endogenous decision and might reflect respondents' assessments that a lower equity share in their portfolio is consistent with their belief changes.

**Figure IV:** Portfolio and Trading Activity

(A) All Respondents

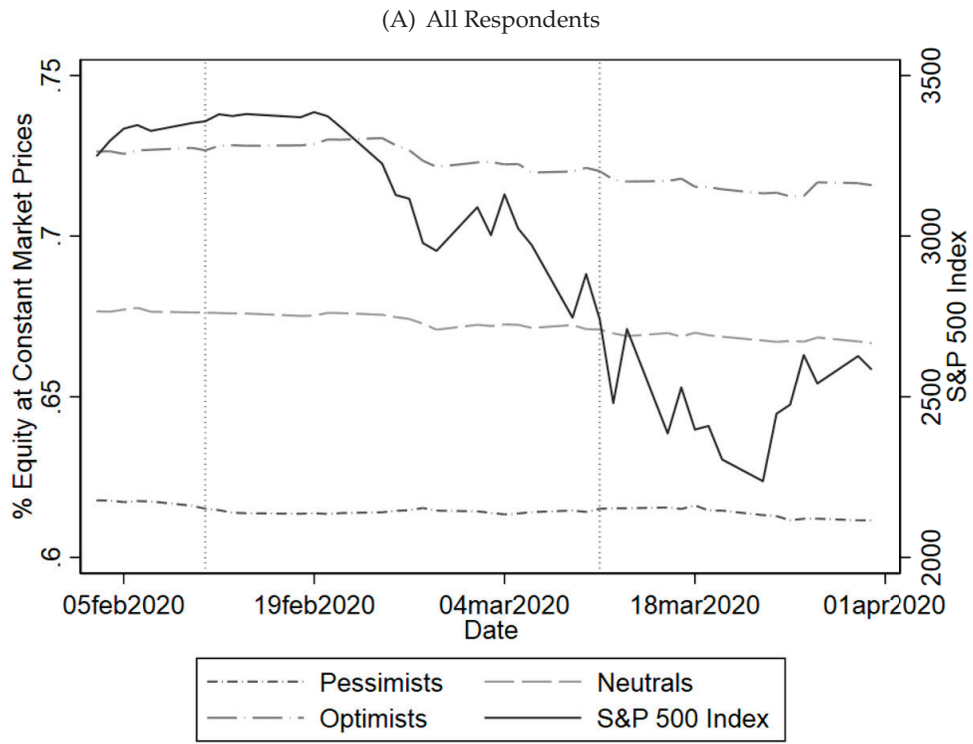

(B) Only Those Respondents Who Traded

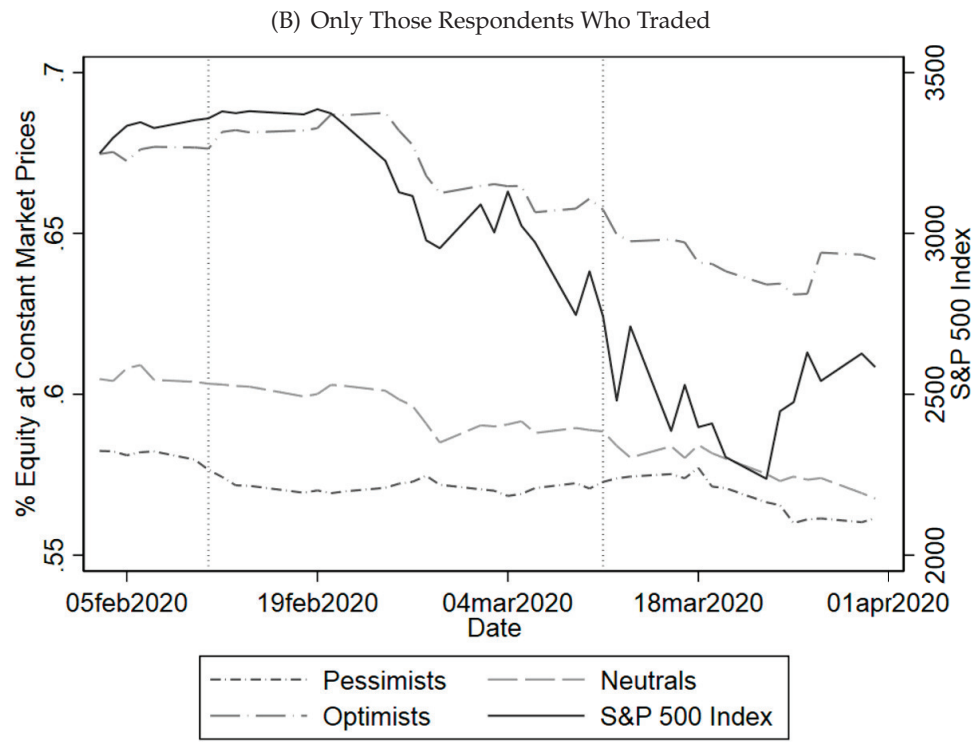

**Note:** Figure shows portfolio dynamics at constant market prices (from January 31st 2020). We group respondents to the February survey by their level of expected 1-year stock market returns, with the optimists being the top tercile, and trace their portfolio activity daily over February and March. Panel A includes all February respondents. Panel B includes only those who actively change their portfolio equity share by at least 1% between January 31st and March 31st. Both panels include for reference the dynamics of the S&P 500 index (right axis).

by the neutrals (active decrease of 0.98%). Initial pessimists had the lowest active change in the equity share, with an active decrease of 0.63%. To inspect this mechanism further, Panel B focuses exclusively on those respondents who actively change their portfolios during the period.[11] Similar patterns appear, but they are now more pronounced, reflecting the fact that a substantial portion of respondents (67% of optimists, 73% of neutrals, and 70% of pessimists) did not change their portfolio during this period. The optimists who trade move their equity percentage from a high of 68% to a low of 64% by the end of March; initial neutrals with active trading move their equity share from 60% to 57%; and initial pessimists from 58% to 56%. There is also an interesting higher-frequency dynamic: the optimists sell their equity during the crash between end of February and early March, and by end of March, after the market rebounds, they buy back part of that equity.

**Correlations across belief changes.** Our final analysis uses our panel data to investigate the joint dynamics of changes in expectations about economic growth and stock market returns across individuals. Table II reports the correlation of individual-level changes in beliefs between the February and March (Panel A) and February and April (Panel B) waves of the survey. For example, the table shows (row 6, column 3) that investors who increased their perceived probability of a stock market disaster also increased the perceived probability of GDP growth disaster.

**Table II:** Correlation Across Belief Changes

| Panel A: February - March | (1) | (2) | (3) | (4) | (5) |
|---|---|---|---|---|---|
| (1) Δ Expected 1Y Stock Return (%) | 1 | | | | |
| (2) Δ Expected 10Y Stock Return (% p.a.) | 0.061 | 1 | | | |
| (3) Δ Prob 1Y Stock Return < -30% (%) | -0.363 | 0.094 | 1 | | |
| (4) Δ Expected 3Y GDP Growth (% p.a.) | 0.155 | 0.140 | -0.063 | 1 | |
| (5) Δ Expected 10Y GDP Growth (% p.a.) | 0.010 | 0.276 | -0.048 | 0.446 | 1 |
| (6) Δ Prob 3Y GDP Growth < -3% p.a. (%) | -0.188 | 0.004 | 0.230 | -0.184 | -0.037 |
| **Panel B: February - April** | (1) | (2) | (3) | (4) | (5) |
| (1) Δ Expected 1Y Stock Return (%) | 1 | | | | |
| (2) Δ Expected 10Y Stock Return (% p.a.) | 0.110 | 1 | | | |
| (3) Δ Prob 1Y Stock Return < -30% (%) | -0.262 | 0.018 | 1 | | |
| (4) Δ Expected 3Y GDP Growth (% p.a.) | 0.213 | 0.125 | -0.154 | 1 | |
| (5) Δ Expected 10Y GDP Growth (% p.a.) | 0.052 | 0.262 | -0.024 | 0.403 | 1 |
| (6) Δ Prob 3Y GDP Growth < -3% p.a. (%) | -0.145 | 0.035 | 0.367 | -0.255 | -0.010 |

**Note:** Table shows cross-sectional correlation of changes in individual beliefs between the February 2020 and March 2020 waves of the GMSU-Vanguard survey (Panel A), and between the February 2020 and April 2020 waves (Panel B).

The first column also highlights that, on average, those investors who became more pessimistic about average stock returns also became more pessimistic about the probability of a stock market crash and a GDP disaster (rows 3 and 6), as well as about the short-run outlook for GDP growth (row 4). However, changes in beliefs about long-run GDP growth and long-run stock market returns (rows 2 and 5) are essentially uncorrelated with changes in short-run expected returns.

---

[11] We include only those respondents to the February survey who actively change their portfolio equity share by at least 1% between January 31st and March 31st.

**Limitations.** Before concluding, we point out a number of possible limitations of the current study. First, like all survey-based studies, the presence of measurement error is a potential concern, especially for the quantitative interpretation of the results. Second, the population of investors we survey is selected both in terms of being Vanguard clients and in terms of choosing to answer the survey. Both of these concerns are extensively discussed in our previous work (Giglio et al., 2019) and here we limit ourselves to pointing out that (i) Vanguard is one of the world's largest asset managers with assets of $6 trillion and over 30 million investors globally (our study draws from the U.S. population of individual retail investors and retirement plan participants, approximately 10 million investors), thus making it an interesting population to study, and (ii) while measurement error and selection are present, we have found our surveys to reveal interesting beliefs that are actually reflected in investors trading decisions. A final concern, more specific to this paper, is that the COVID-19 crisis is a particular shock with a number of idiosyncratic components. It is therefore unclear how many of the patterns here might generalize to other large economic shocks. It suffices to say that shocks of this magnitude are so rare that some advancement in our understanding can be achieved by their study even after considering their idiosyncratic limitations.

## III   Takeaways for Economic Theory

Our purpose in this paper is to document novel patterns of belief dynamics and trading activity during a substantial market crash. These patterns represent new data moments that can be useful to calibrate and evaluate economic models. In this section, we briefly highlight the main qualitative implications for various models, while leaving a quantitative exploration that requires more theoretical structure to future research.

Our data is perhaps most directly related to the rare-disaster model of macro-finance (Rietz, 1988; Barro, 2006). Our data supports an interesting feature of versions of this model with time-varying disaster probabilities (Gabaix, 2012; Wachter, 2013): that the occurrence of a crash is associated with higher (perceived) probability of future disasters. This is exactly what we see in the data. However, these models also imply that, precisely because the probability of disasters increases, expected returns should also increase following a stock market crash.[12] This latter prediction is not supported by our data. The prediction of higher expected returns following a stock market crash is a shared feature of many rational-expectation asset pricing models, and in this sense, the empirical failure is common across this class of models.[13] Our paper offers a useful

---

[12]More precisely, risk premia should increase. In these models, like in the data, the risk-free rate decreases with the crash. In our data expected returns decreased from 6.37% to 1.3%, while short-term rates (1 year treasury bills) decreased from 1.44% (February 11th) to 0.39% (March 11th). We conclude that expected excess returns (a form of risk premium) decreased in the data with the crash.

[13]The literature's focus on representative agent models makes the mapping with our data difficult in the absence of an explicit aggregation theorem. Nonetheless, the literature that documented the negative correlation between expected rational returns and those elicited via surveys routinely equates average survey-beliefs to those of representative agents (see Greenwood and Shleifer, 2014). On exception is Chen, Joslin and Tran (2012), who explicitly model heterogenous beliefs about time-varying rare disasters. In that model, the optimists underestimate the probability of a disaster and, therefore, expect higher returns than the pessimists. When the probability of a disaster increases expected returns for both types of agents increase. In the data, with the exception of a portion of the most pessimistic investors, all investors lower their expected returns after the crash.

testing ground for future evolutions of these theories.

Our data also highlight mechanisms that are at the core of the theoretical literature on heterogenous beliefs and trading (Harrison and Kreps, 1978; Scheinkman and Xiong, 2003; Simsek, 2013). Ex-ante optimists have more exposure to equity than ex-ante pessimists and, as a result, they lose more wealth when the crash occurs. In those models, changes of beliefs play a crucial role in generating trading activity when investor beliefs "cross" each other. Consistent with such mechanisms, we find that ex-ante optimists lower their beliefs the most after the crash, and correspondingly sell the most equity. In most of the literature, belief changes are idiosyncratic at the individual level, but we show in Table I that even at the individual level there are substantial belief changes associated with heterogenous reactions to aggregate events. A full quantitative evaluation is outside the scope of this paper and would probably require enriching the existing models with additional channels, but our analysis suggests that this class of models can be a promising direction to explain the patterns of beliefs and trading that we document.

To conclude, our study provides a unique real-time look inside the mind of stock market participants during the COVID-19 crisis, and the associated stock market crash. It shows that investors turned more pessimistic and increased their perceived probabilities of catastrophic events in terms of real economic outcomes and further stock market declines. We also find that investors reduced their equity exposures according to changes in their expectations. At the same time, we find that investors also formed a nuanced view of long term prospects. Short-term pessimism was matched with unchanged or even improved long-run expectations. By documenting these dynamics of beliefs and trading during a large market crash, and by characterizing their heterogeneity across investors, we hope to bring useful additional moments that can help test ad calibrate macro-finance theories.

## A.1 INVITATION EMAIL AND SURVEY FLOW

In this Appendix, we present screenshots of one complete survey flow. In this iteration of the flow, questions about expected stock returns were asked ahead of questions about expected GDP growth; the survey implementation randomizes across these two blocks of questions. We begin by reviewing the invitation email sent to individuals from Vanguard.

| | |
|---|---|
| **Subject:** | [TEST]We need your help, Jane Doe |
| **From:** | Vanguard (vanguard@eonline.e-vanguard.com) |
| **To:** | oea_test@yahoo.com; |
| **Date:** | Monday, February 13, 2017 10:58 AM |

**Vanguard**

### Vanguard would like your input

Dear Jane Doe:

Vanguard is conducting a study to understand how investors are thinking about the future of the stock market, the economy and interest rates.

We are inviting you to provide us with your thoughts by completing a short survey. This survey should take less than ten minutes to complete.

This survey is not a test of your knowledge. Rather, it asks only about your beliefs and expectations. Importantly, it does not ask for any personal financial information.

The results of the survey will be used for research purposes only. This survey is not sales-related in any way. Your responses will be reported in aggregate with other responses. We plan to publish the results in an article or research report on vanguard.com.

To participate in the survey, please click here.

**Take the survey**

We'd also like to send you this survey up to six times in the coming year, to see if your beliefs are changing. If you want to be removed from this study, you have the option to click the unsubscribe link below.

If you have any questions about this survey, please call **800-662-2739** and refer to this code: **EXP**.

Thank you for participating, and for sharing your thoughts with Vanguard.

Regards,

Stephen Utkus
Principal
Vanguard

**Legal notices**
Please click here to be removed from this study.

© 2017 The Vanguard Group, Inc. All rights reserved. Privacy policy

455 Devon Park Drive | Wayne, PA 19087-1815 | vanguard.com
EXP

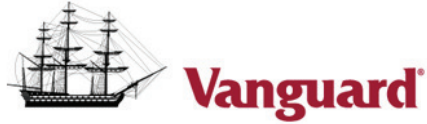

Dear Investor,

Thank you for participating in this study.

At Vanguard we are interested in understanding investor views on the future of the stock market, the economy and interest rates. We plan to create an investor sentiment index to share these findings with the investing public.

This is a short survey that should take you no more than 5-10 minutes to complete.

The survey does not collect any personal information. It relies on your general knowledge.

**If you feel you are not familiar with a topic, that is fine. Please just give us your best prediction.**

**Please do not use the browser's navigation button to move through the survey.**

Next

---

For these questions, we would like to know what you are expecting the future returns of the US stock market to be.

Next

---

What do you expect the return of the US stock market to be **over the next 12 months**?

Note: This expected return is the change in value, in percent, that you expect to receive **over the next 12 months** from investing in a portfolio that holds all stocks listed on the US stock market. It includes both dividends and capital gains/losses.

*(Please answer only with a positive or negative numeric value, with at most 1 decimal.)*

[____] % over the next 12 months

Next

**radius** GLOBAL MARKET RESEARCH

---

What do you expect the **average** annual return of the US stock market to be **over the next 10 years**?

Note: This expected return is the change in value, in percent, that you expect to receive **each year on average over the next 10 years** from investing in a portfolio that holds all stocks listed on the US stock market. It includes both dividends and capital gains/losses.

*(Please answer only with a positive or negative numeric value, with at most 1 decimal.)*

[ ______ ] % per year, over the next 10 years

**Next**

---

**radius**

---

In this question we present you with five possible scenarios for US stock market returns **over the next 12 months**:

The US stock market return will be...

- Scenario 1: **more than 40%** over the next year.
- Scenario 2: **between 30% and 40%** over the next year.
- Scenario 3: **between -10% and 30%** over the next year.
- Scenario 4: **between -30% and -10%** over the next year.
- Scenario 5: **less than -30%** over the next year.

Please let us know how likely you think it is that each scenario will occur.

Please type in the number to indicate the probability, in percent, that you attach to each scenario. The probabilities of the five scenarios have to sum up to 100%. The graphic bar chart on the right updates automatically to reflect your answers.

*(Please answer only with a positive numeric value, with at most 1 decimal.)*

| | | |
|---|---|---|
| more than 40% | [ ___ ] % | |
| between 30% and 40% | [ ___ ] % | 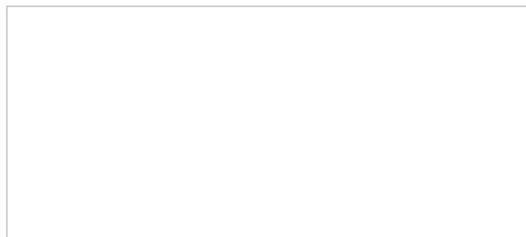 |
| between -10% and 30% | [ ___ ] % | |
| between -30% and -10% | [ ___ ] % | |
| less than -30% | [ ___ ] % | |
| Total | 0.0% | |

**Remaining probability to fill in: 100.0%**

**Next**
Powered by Confirmit
A.3A.3

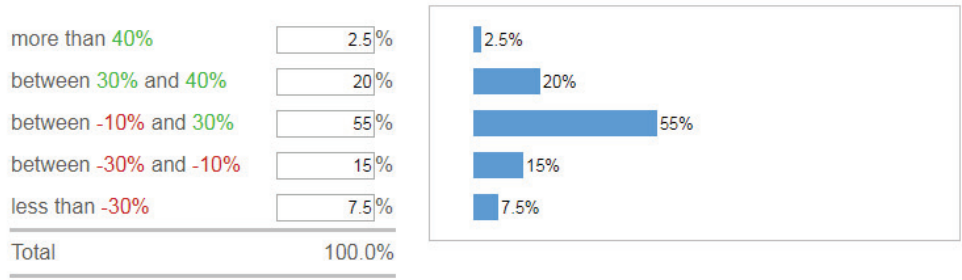

**GLOBAL MARKET RESEARCH**
radius

How confident are you with your answers to the questions about the stock market that you were just asked?

- ○ Extremely confident
- ○ Very confident
- ○ Somewhat confident
- ○ Not very confident
- ○ Not at all confident

[Next]

**GLOBAL MARKET RESEARCH**
radius

In the next questions, we would like to know what you are expecting future economic growth in the US to be.

**Again, even if you feel that you are not familiar with the topic, please give us your best prediction.**

[Next]

**GLOBAL MARKET RESEARCH**
radius

What do you expect the **average** annual growth rate of real GDP in the US to be **over the next 3 years**?

Note: Real Gross Domestic Product (GDP) is a measure of economic activity. Real GDP is the total real value of goods and services produced in the US in a year.

*(Please answer only with a positive or negative numeric value with at most 1 decimal.)*

[____] % per year, over the next 3 years

[Next]

![radius - GLOBAL MARKET RESEARCH]

What do you expect the **average** annual growth rate of real GDP in the US to be **over the next 10 years**?

Note: Real Gross Domestic Product (GDP) is a measure of economic activity. Real GDP is the total real value of goods and services produced in the US in a year.

*(Please answer only with a positive or negative numeric value with at most 1 decimal.)*

[____] % per year, over the next 10 years

**Next**

---

![radius]

In this question we present you with five possible scenarios for US real GDP **average annual growth rate, over the next 3 years**:

US real GDP average annual growth rate over the next 3 years will be...

- Scenario 1: **more than 9%** per year.
- Scenario 2: **between 3% and 9%** per year.
- Scenario 3: **between 0% and 3%** per year.
- Scenario 4: **between -3% and 0%** per year.
- Scenario 5: **less than -3%** per year.

Please let us know how likely you think it is that each scenario will occur.

Please type in the number to indicate the probability, in percent, that you attach to each scenario. The probabilities of the five scenarios have to sum up to 100%. The graphic bar chart on the right updates automatically to reflect your answers.

*(Please answer only with a positive numeric value, with at most 1 decimal.)*

| | | |
|---|---|---|
| more than 9% | [____] % | |
| between 3% and 9% | [____] % | |
| between 0% and 3% | [____] % | |
| between -3% and 0% | [____] % | |
| less than -3% | [____] % | |
| Total | 0.0% | |

**Remaining probability to fill in: 100.0%**

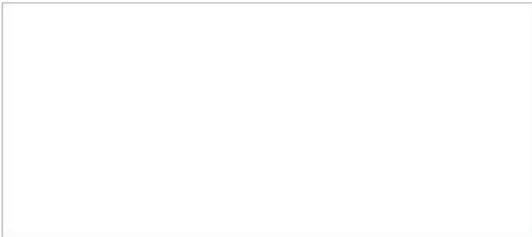

**Next**

*radius* GLOBAL MARKET RESEARCH

How difficult were the questions about real GDP growth that you were just asked?

- Not at all difficult
- Not very difficult
- Somewhat difficult
- Very difficult
- Extremely difficult

[Next]

Powered by Confirmit

---

*radius* GLOBAL MARKET RESEARCH

How confident are you with your answers to the questions about real GDP growth that you were just asked?

- Extremely confident
- Very confident
- Somewhat confident
- Not very confident
- Not at all confident

[Next]

Powered by Confirmit

---

*radius* GLOBAL MARKET RESEARCH

In these final questions, we would like to know what you are expecting future returns on US bonds and future US interest rates to be.

**Again, even if you feel that you are not familiar with the topic, please give us your best prediction.**

[Next]

Powered by Confirmit

Suppose that you were to buy a 10-year US Treasury bond today that makes all of its payments at maturity 10 years from now.

Suppose that you were to sell this bond a year from today. What do you expect the return from this bond investment to be **over the next 12 months**?

Note: This expected return is the change in price of the bond that you expect to occur during the next 12 months.

*(Please answer only with a positive or negative numeric value with at most 1 decimal.)*

[        ] % over the next 12 months

Next

How difficult were the questions about bonds and interest rates that you were just asked?

○ Not at all difficult
○ Not very difficult
○ Somewhat difficult
○ Very difficult
○ Extremely difficult

Next

How confident are you with your answers to the questions about bonds and interest rates that you were just asked?

○ Extremely confident
○ Very confident
○ Somewhat confident
○ Not very confident
○ Not at all confident

Next